\documentstyle[aps,preprint,epsfig]{revtex}

\begin{document}
\draft

\title{Epidemic spreading in scale-free networks}
 
\author{Romualdo Pastor-Satorras$^1$
and Alessandro Vespignani$^2$}

\address{$^1$ Dept. de F\'{\i}sica i Enginyeria Nuclear,
  Universitat Polit\`{e}cnica de Catalunya,\\
  Campus Nord, Bloc B4/B5,  08034 Barcelona, Spain   \\
  $^2$ The Abdus Salam International Centre 
  for Theoretical Physics (ICTP),
  P.O. Box 586, 34100 Trieste, Italy
}
\date{\today}
\maketitle

\begin{abstract}
  The Internet, as well as many other networks, has a very complex
  connectivity recently modeled by the class of scale-free networks.
  This feature, which appears to be very efficient for a
  communications network, favors at the same time the spreading of
  computer viruses.  We analyze real data from computer virus
  infections and find the average lifetime and prevalence of viral
  strains on the Internet. We define a dynamical model for the
  spreading of infections on scale-free networks, finding the absence 
  of an epidemic threshold and its associated critical behavior.
  This new epidemiological framework rationalize data of computer 
  viruses and could help in the understanding of other spreading 
  phenomena on communication and social networks.
\end{abstract}


Many social, biological, and communication systems can be properly
described by complex networks whose nodes represent individuals or
organizations, and links mimic the interactions among
them\cite{net1,amaral}.  Particularly interesting examples are the
Internet and the world-wide-web, which have been extensively studied
because of their technological and economical
relevance\cite{faloutsos,int,www99}. These studies have revealed,
among other facts, the {\em scale-free nature} of these networks
\cite{faloutsos,www99}. This results in the power-law distribution
$P(k)\sim k^{-\gamma}$ for the probability that a node of the network
has $k$ connections to other nodes, with an exponent $\gamma$ that
ranges between $2$ and $3$. The importance of local clustering is
indeed the key ingredient in the modeling of these networks with the
recent introduction of scale-free (SF) graphs \cite{barab99}.

In view of the wide occurrence of complex networks in nature it is of
great interest to inspect the effect of their features on epidemic and
disease spreading \cite{mn99}, and more in general in the context of
the nonequilibrium phase transitions typical of these 
phenomena \cite{marro99}. The
study of epidemics on these networks finds an immediate
practical application in the understanding of computer virus spreading
\cite{virus,scientific97}, and could also be relevant to the fields
of epidemiology \cite{biomodels} and pollution control \cite{pol}.

In this Letter, we
analyze data from real computer virus epidemics, providing a
statistical characterization that points out the importance of
incorporating the peculiar topology of scale-free networks in the
theoretical description of these infections. With this aim, we study
by large scale simulations and analytical methods  the
susceptible-infected-susceptible \cite{biomodels} model on 
SF graphs.  We find the absence of an epidemic threshold and its
associated critical behavior, which implies that SF networks  are
prone to the spreading and the  persistence of infections whatever
spreading rate the epidemic agents possess. The absence of the
epidemic threshold---a standard element in mathematical
epidemiology\cite{biomodels}--radically changes 
many of the standard conclusions drawn
in epidemic modeling. The present results are also relevant in the field
of absorbing-state phase transitions and catalytic reactions\cite{marro99}.

The epidemiological analysis of computer viruses has been the subject
of a continuous interest in the computer science community
\cite{scientific97,ieee93,murray88,kep91}, following mainly approaches
borrowed from biological epidemiology~\cite{biomodels}.  The standard
model used in the study of computer virus infections is the
susceptible-infected-susceptible (SIS) epidemiological model. Each
node of the network represents an individual and each link is a
connection along which the infection can spread to other systems.
This model relies on a coarse grained description of individuals in
the population.  Individuals exist only in two discrete states,
``healthy'' or ``infected''.  At each time step, each susceptible
(healthy) node is infected with rate $\nu$ if it is connected to one
or more infected nodes. At the same time, infected nodes are cured and
become again susceptible with rate $\delta$, defining an effective
spreading rate $\lambda=\nu/\delta$.  Without lack of generality, we
can set $\delta=1$. The updating can be performed both with parallel
and sequential dynamics\cite{marro99}.  In models with local
connectivity (Euclidean lattices and mean-field models), the most
significant result is the general prediction of a nonzero epidemic
threshold $\lambda_c$~\cite{marro99,biomodels}.  If the value of
$\lambda$ is above the threshold, $\lambda\geq \lambda_c$, the
infection spreads and becomes persistent. Below it, $\lambda <
\lambda_c$, the infection dies out exponentially fast.  The epidemic
threshold is actually equivalent to a critical point in a
nonequilibrium phase transition. In this case, the critical point
separates an active phase with a stationary density of infected nodes
from a phase with only healthy nodes and null activity.  In
particular, it is easy to recognize that the SIS model is a
generalization of the contact process (CP) model, that has been
extensively studied in the context of absorbing-state phase
transitions \cite{marro99}.  Statistical observations of virus
incidents in the wild, on the other hand, indicate that all viruses
that are able to pervade, spread much slower than exponentially, and
saturate to a very low level of persistence, affecting just a tiny
fraction of the total number of computers \cite{scientific97}.  This
fact is in striking contradiction with the theoretical predictions
unless in the very unlikely chance that {\em all} computer viruses
have an effective spreading rate tuned just infinitesimally above the
threshold. This points out that the view obtained so far with the
modeling of computer virus epidemics is very instructive but not
completely adequate to represent the real phenomenon.

In order to gain further insight on the spreading properties of
viruses in the wild, we have analyzed the prevalence data reported by
the Virus Bulletin~\cite{bull00} from February 1996 to March 2000.  We
have analyzed in particular the {\em surviving probability} of
homogeneous groups of viruses, classified according to their infection
mechanism \cite{virus}.  We consider the total number of viruses of a
given strain that born and die within our observation window.  Hence,
we calculate the surviving probability $P_s(t)$ of the strain as the
ratio of viruses still alive at time $t$ after their birth and the
total number of observed viruses. Fig. 1 shows that the surviving
probability suffers a sharp drop in the first two months of a virus'
life. This is a well-known feature~\cite{scientific97,ieee93}
indicating that statistically only a small percentage of viruses give
rise to a significant outbreak in the computer community.  Fig. 1, on
the other hand, shows for larger times a clean exponential tail,
$P_s(t)\sim\exp (-t/\tau)$, where $\tau$ represents the characteristic
life-time of the virus strain~\cite{nota3}.  The numerical fit of the
data yields $\tau\simeq 14$~months for boot and macro viruses and
$\tau\simeq 6-9$ months for file viruses. These characteristic times
are impressively large if compared with the interval in which
anti-virus software is available on the market (usually within days or
weeks after the first incident report) and indicate that the viral
persistence time scale is more related to the implementation of
prophylactic safety measures than to the timely availability of the
specific anti-virus. These external factors, however, are not possibly
competing on the short time scale of the viruses spread (days or
weeks), and again we face the very puzzling question of why viruses
seem to have access to persistent low prevalence levels but never grow
exponentially.

The key point to understand the puzzling properties exhibited by
computer viruses resides in the capacity of many of them to be borne
by electronic mail as an apparently innocuous attachment
\cite{scientific97}. Having this property in mind, it is easy to
realize that the topology of the connections between individuals
cannot be correctly represented by an Euclidean lattice, or a
mean-field model. In this sense, these connections should instead have
essentially the topology of the Internet, through which electronic
mail travels. The scale-free connectivity of the Internet implies that
each node has a statistically significant probability of having a very
large number of connections compared to the average connectivity
$\left<k\right>$ of the network.  That opposes to conventional random
networks (local or nonlocal) in which each node has approximately the
same number of links $k\simeq \left<k\right>$ \cite{exponential}. The
fact that all virus strains qualitatively show the same statistical
features indicates that very likely all of them spread on networks
with connectivity properties analogous to those of the
Internet~\cite{nota5}. It is then natural to foresee that scale-free
properties should be included in a theory of epidemic spreading of
computer viruses.

To address the effects of scale-free connectivity in epidemic
spreading we study the SIS model on the SF network. We consider the
graph generated by using the algorithm devised in
Refs.~\cite{barab99}: We start from a small number $m_0$ of
disconnected nodes; every time step a new vertex is added, with $m$
links that are connected to an old node $i$ with probability
$k_i/\sum_j k_j$. After iterating this scheme a sufficient number of
times, we obtain a network composed by $N$ nodes with connectivity
distribution $P(k) \sim k^{-3}$ and average connectivity $\left<k
\right> = 2 m$. In this work we take $m=3$.  We have performed
numerical simulations on graphs with number of nodes ranging from
$N=10^3$ to $N=8.5\times 10^6$ and studied the variation in time and
the stationary properties of the density of infected nodes $\rho$ in
surviving infections; i.e. the virus prevalence. Initially we infect
half of the nodes in the network, and iterate the rules of the SIS
model with parallel updating. After an initial transient regime, the
system stabilizes in a steady state with a constant average density of
infected nodes.  The prevalence is computed averaging over at least
$100$ different starting configurations, performed on at least $10$
different realizations of the random networks.

The first arresting evidence from simulations is the {\em absence} of
an epidemic threshold, i.e., $\lambda_c=0$. In Fig. 2 we show the
virus prevalence in the steady state that decays with decreasing
$\lambda$ as $\rho\sim \exp(-C/\lambda)$, where $C$ is a constant.
This implies that for any finite value of $\lambda$ the virus can
pervade the system with a finite prevalence, in sufficiently large
networks.  In all networks with bounded connectivity the steady state
prevalence is always null below the epidemic threshold; i.e. all
infections die out.  Further evidence to our results is given by the
total absence of scaling of $\rho$ with the number of nodes that is,
on the contrary, typical of epidemic transitions in the proximity of a
finite threshold \cite{marro99}. This allows us to exclude the
presence of any spurious results due to network finite size effects.
The present result can be intuitively understood by noticing that for
usual lattices, the higher the node's connectivity, the smaller the
epidemic threshold. In a SF network the unbounded fluctuations in
connectivity ($\left< k^2 \right> = \infty$) plays the role of an
infinite connectivity, annulling thus the threshold.

Finally, we analyze the spreading of infections starting from a
localized virus source.  We observe that the spreading
growth in time has an algebraic form, that is in agreement with real
data that never found an exponential increase of a virus in the wild.
Noteworthy, by applying the definition of surviving probability
$P_s(t)$ used to analyze real data, we recover in our model the same
exponential behavior in time (see Fig. 3a). The characteristic life-time
depends on the spreading rate and the network sizes, allowing us to
relate the average lifetime of a viral strain with an effective
spreading rate and the Internet size\cite{notesize}.  
At the same time, the divergence of lifetimes for
larger networks points out the possible increasing of the viruses
lifetime during the eventual expansion of the Internet.

We can also approach the system analytically by writing the mean-field
equation governing the time evolution of $\rho(t)$. In order to take into
account connectivity fluctuations, we consider the relative density
$\rho_k(t)$ of infected nodes with given connectivity $k$; i.e the
probability that a node with $k$ links is infected. 
The dynamical mean-field reaction rate equations 
can be written as \cite{marro99}
\begin{equation}
  \partial_t \rho_k(t) = -\rho_k(t) +\lambda k (1-\rho_k(t))\Theta(\lambda).
\end{equation}
The creation term considers the probability that a node with $k$ links
is healthy $(1-\rho_k(t))$ and gets the infection via a connected node.
The probability of  this  event is proportional to the infection
rate, the number of connections, and the probability $\Theta(\lambda)$
that any given link points to an infected node.  
By imposing stationarity ($\partial_t \rho_k(t)=0$) 
we find the stationary densities 
\begin{equation}
  \rho_k=\frac{k \lambda\Theta(\lambda)}{1+k \lambda\Theta(\lambda)},
  \label{dep}
\end{equation}
denoting that the higher the node connectivity, the higher the
probability to be infected. This inhomogeneity must be taken into
account in the self-consistent calculation of $\Theta(\lambda)$. Indeed,
the probability that a link points to a node with $s$ links is
proportional to $sP(s)$.  In other words, a randomly chosen link is
more likely to be connected to a node with high connectivity, yielding
\begin{equation}
  \Theta(\lambda)=\sum_k \frac{kP(k)\rho_k}{\sum_s sP(s)}.
\label{first}
\end{equation}
Since $\rho_k$ is on its turn function of $\Theta(\lambda)$, we obtain
a consistency equation that allows to to find $\Theta(\lambda)$.
Finally we can evaluate the behavior of $\rho$ by solving the second
consistency relation
\begin{equation}
  \rho=\sum_kP(k)\rho_k,
  \label{second}
\end{equation}
that expresses the average density of infected nodes in the system.
In the SF model considered here, 
we have a connectivity distribution $P(k)=2m^2/k^{-3}$, 
where $k$ as approximated as a continuous variable
\cite{barab99}. In this case, integration of Eq.(\ref{first}) allows
to write $\Theta(\lambda)$ as
\begin{equation}
  \Theta(\lambda)=\frac{e^{-1/m\lambda}}{\lambda
  m}(1-e^{-1/m\lambda})^{-1},
\end{equation}
from which, using Eq.(\ref{second}), we find at lowest order in
$\lambda$: 
\begin{equation}
  \rho=2e^{-1/m\lambda} + h.o.~.
\end{equation}
This very intuitive calculation recovers the numerical findings and
confirms the surprising absence of any epidemic threshold or critical
point in the model; i.e $\lambda_c=0$. Finally, as a further check of
our analytical results, we have numerically computed in our model the
relative densities $\rho_k$ , recovering the predicted dependence upon
$k$ of Eq.(\ref{dep}) (see Fig. 3b). It is also worth remarking that the
present framework can be generalized to networks with $2<\gamma\leq
3$, recovering qualitatively the same results~\cite{psv}.

The emerging picture for epidemic spreading in complex networks
emphasizes the role of topology in epidemic modeling. In particular,
the absence of epidemic threshold and critical behavior in a wide
range of scale-free network provide an unexpected result that changes
radically many standard conclusions on epidemic spreading.  This
indicates that infections can proliferate on these scale-free networks
whatever spreading rates they may have. These very bad news are,
however, balanced by the exponentially small prevalence for a wide
range of spreading rates ($\lambda<<1$).  This point appears to be
particularly relevant in the case of technological networks such as
the Internet and the world-wide-web that show scale-free connectivity
with exponents $\gamma\simeq 2.5$ \cite{int,www99}. For instance, the
present picture fits perfectly with the observation from real data of
computer virus spreading, and could solve the long standing problem of
the generalized low prevalence of computer viruses without assuming
any global tuning of the spreading rates. The peculiar
properties of scale-free networks also open the path to many other
questions concerning the effect of immunity and other modifications of
epidemic models. As well, the critical properties of many
nonequilibrium systems could be affected by the topology of scale-free
networks. Given the wide context in which scale-free networks appears,
the results obtained here and the proposed investigations could have
intriguing implications in biology and social systems.

This work has been partially supported by the European Network 
Contract No. ERBFMRXCT980183.  RP-S also acknowledges support from the
grant CICYT PB97-0693. We thank S.~Franz, M.-C. Miguel, R. V. Sol\'e,
M.~Vergassola, S.~Zapperi and R. Zecchina for helpful comments and
discussions.

\newpage

\begin{figure}[t]
  \centerline{\epsfig{file=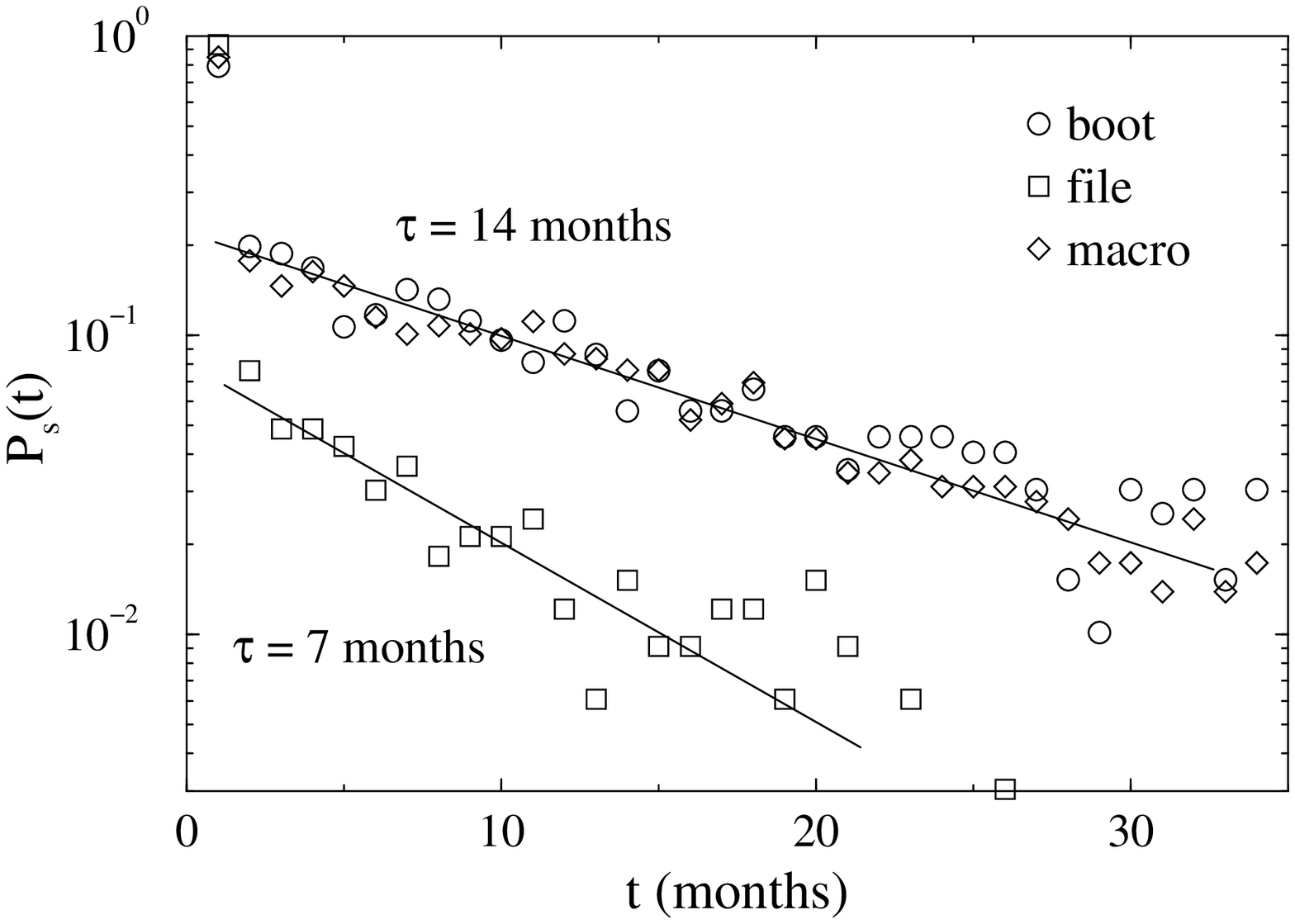, width=13cm}}
  \caption{Surviving probability for viruses in the wild. The 814
    different viruses analyzed have been grouped in three main strains
    \protect\cite{virus}: file viruses infect a computer when running
    an infected application; boot viruses also spread via infected
    applications, but copy themselves into the boot sector of the
    hard-drive and are thus immune to a computer reboot; macro viruses
    infect data files and are thus platform-independent.  It is
    evident in the plot the presence of an exponential decay, with
    characteristic time $\tau \simeq 14$ months for macro and boot
    viruses and $\tau \simeq 7$ months for file viruses.}
\end{figure}

\newpage

\begin{figure}[t]

  \centerline{\epsfig{file=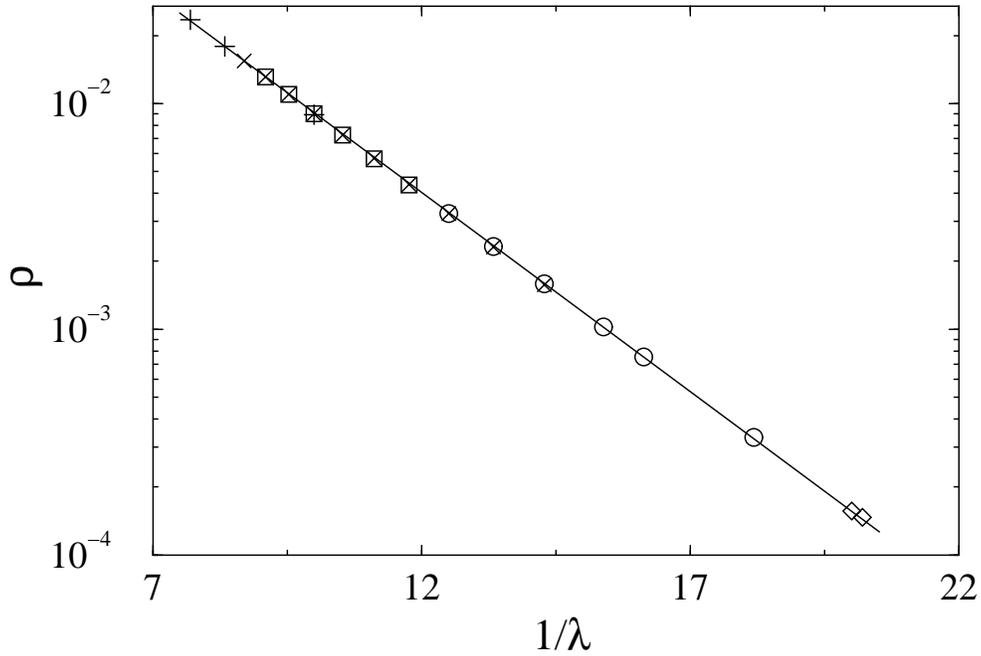, width=13cm}}
  
  \caption{Persistence $\rho$ as a function of $1/\lambda$ for
    different network sizes: $N=10^5$ ($+$), $N=5 \times 10^5$
    ($\Box$), $N=10^6$ ($\times$), $N=5 \times 10^6$ ($\circ$), and
    $N=8.5 \times 10^6$ ($\Diamond$). The linear behavior on the
    semi-logarithmic scale proves the stretched exponential behavior
    predicted for $\rho$. The full line is a fit to the form $\rho
    \sim\exp(-C/\lambda)$.}
\end{figure}

\newpage

\begin{figure}[t]
  \centerline{\epsfig{file=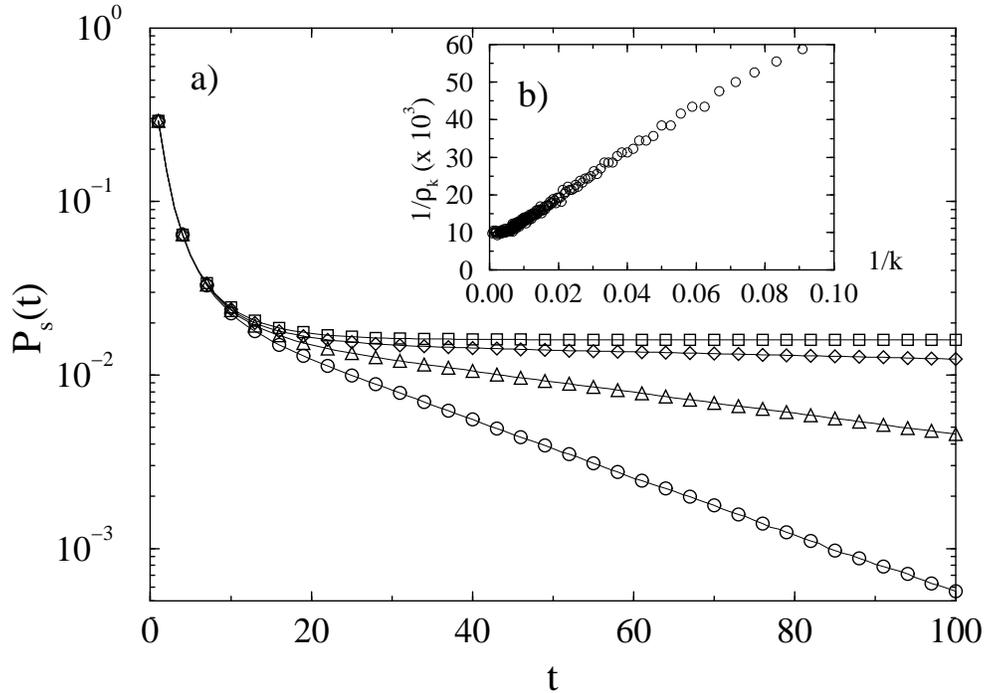, width=13cm}}
  \caption{a) Surviving probability
    $P_s(t)$ for a spreading rate $\lambda=0.065$ in scale-free
    networks of size $N=5 \times 10^5$ ($\Box$), $N=2.5 \times 10^4$
    ($\Diamond$), $N=1.25 \times 10^4$ ($\triangle$), and $N=6.25
    \times 10^3$ ($\circ$). The exponential behavior, following a
    sharp initial drop, is compatible with the data analysis of Fig.
    1.  b) Relative density $\rho_k$ versus $k^{-1}$ in a SF 
    network of size $N=5 \times 10^5$ and spreading 
    rate $\lambda=0.1$.  The plot recovers
    the form predicted in Eq.(\protect\ref{dep}).}
\end{figure}


\begin{thebibliography}{10}
\bibitem{nota3} This is the usual way in which it is determined the
  survival probability in numerical simulations of spreading
  models, ref. \cite{marro99}.
\bibitem{nota5} For example, the friendship network of people with
  which one is ready to exchange infected floppy disks.
\bibitem{notesize}
        This characteristic  scaling is often encountered at 
        absorbing-state phase transitions in finite size 
        systems\cite{marro99}. In general, $P_s(\infty)$ is 
        finite only for infinite size networks.
\end{thebibliography}
\end{document}